\documentclass[conference]{IEEEtran}
\IEEEoverridecommandlockouts

\usepackage{cite}
\usepackage{graphicx}
\usepackage{amsmath}
\usepackage{amssymb}
\usepackage{booktabs}
\usepackage{xcolor}
\usepackage{tikz}
\usetikzlibrary{arrows.meta,positioning,fit,backgrounds,calc}
\usepackage[caption=false,font=footnotesize]{subfig}
\usepackage{url}
\usepackage{tikz}
\usepackage{quantikz}
\usepackage{listings}
\usepackage{hyperref}

\graphicspath{{figures/}}

\definecolor{codekw}{rgb}{0.15,0.15,0.7}
\definecolor{codecm}{rgb}{0.35,0.55,0.35}
\lstdefinestyle{dqc}{
  language=Python, basicstyle=\ttfamily\scriptsize, keywordstyle=\color{codekw},
  commentstyle=\color{codecm}\itshape, showstringspaces=false, columns=fullflexible,
  breaklines=true, aboveskip=3pt, belowskip=1pt, xleftmargin=2pt,
}

\newcommand{\dqc}{\texttt{SeQUeNCe\hspace{-0.15em}.\hspace{-0.15em}DQC}}
\newcommand{\telegate}{\textsc{TeleGate}}
\newcommand{\teledata}{\textsc{TeleData}}

\newcommand{\para}[1]{\smallskip \noindent\textbf{#1}}

\begin{document}
\bstctlcite{IEEEexample:BSTcontrol}

\title{\dqc: Full-stack Simulation of Distributed Quantum Computing}

\author{
  \IEEEauthorblockN{Sanjay Suresh$^*$, Caitao Zhan$^{\dagger}$, Joaquin Chung$^{\dagger}$, Rajkumar Kettimuthu$^{\dagger}$}
  \IEEEauthorblockA{$^{*}$University of Wisconsin-Madison, $^{\dagger}$Argonne National Laboratory (USA)}
  \IEEEauthorblockA{ssuresh27@wisc.edu, czhan@anl.gov, chungmiranda@anl.gov, kettimuthu@anl.gov}
 
 \thanks{This material is based upon work supported by the U.S.~Department of Energy Office of Science National Quantum Information Science Research Centers as part of the Q-NEXT Center.}
 
 }

\maketitle


\begin{abstract}
Distributed quantum computing (DQC) connects multiple quantum processing units to overcome the scale limits of monolithic hardware. 
We introduce \dqc, a new top-level module within the SeQUeNCe package that compiles monolithic circuits into topology-aware execution plans and coordinates their execution using SeQUeNCe's existing quantum-network modules and protocols. 
We validate \dqc{} by reproducing a two-node trapped-ion Grover experiment in simulation and comparing success probabilities with the reported 71\% benchmark.
We then evaluate how compiler strategy, topology, and memory capacity jointly govern distributed circuit execution time and success rate. 
In these experiments, our hybrid compilers successfully identified mixed \telegate{}/\teledata{} execution plans that outperformed pure gate-teleportation strategies.
By capturing the physical overhead of distributed quantum primitives, \dqc{} provides a rigorous platform for co-designing DQC software and hardware architectures.

\end{abstract}
\section{Introduction}
\label{sec:introduction}

By interconnecting multiple quantum processors, distributed quantum computing (DQC)~\cite{calaffi-dqc-cn24,dqc-review-2025} offers a promising path beyond the scale limits of monolithic hardware toward applications such as quantum chemistry~\cite{vqe-tqe21}, blind quantum computing~\cite{blind-science25}, large-integer factorization~\cite{factor-prr26}, and distributed unstructured search~\cite{grovers-nature25}.
Realizing DQC systems, however, remains exceptionally challenging: it requires quantum processors capable of distributing high-fidelity entanglement at practical rates~\cite{neutral-atom-26} over reliable quantum network channels and protocols~\cite{single-heralded-pra1999,barrett-kok-pra05,acp-qcnc25,interqnet-tqe-2026}. 
Because building and testing physical testbeds are costly and difficult~\cite{grovers-nature25}, software \emph{simulation} serves as a vital approach for designing, evaluating, and optimizing DQC architectures, algorithms, and protocols.

Motivated by this need, numerous efforts to simulate DQC have emerged~\cite{interlin-q-2021,sendq-sc21,dqc-execution-hpqci24,sreraman-dqc-2025,cunqa-2025,simdisq-2025,q-dice-2026,switchqnet-isca25,dqc_simulator-2026}. 
Interlin-q~\cite{interlin-q-2021} and SENDQ~\cite{sendq-sc21} operate primarily at the circuit level using simplified network abstractions without considering noise.
More recent works incorporate hardware noise and focus on job scheduling~\cite{dqc-execution-hpqci24}, dynamic circuits~\cite{sreraman-dqc-2025}, or HPC-oriented emulation~\cite{cunqa-2025}.
SimDisQ~\cite{simdisq-2025} and Q-DICE~\cite{q-dice-2026} further consider inter- and intra-QPU topologies, while SwitchQNet~\cite{switchqnet-isca25} provides network-aware compilation and communication scheduling for switch-based quantum data centers.
The dqc\_simulator~\cite{dqc_simulator-2026}, built on NetSquid, places greater emphasis on quantum-network simulation. 

A central challenge is understanding how compilation choices and quantum network behavior jointly determine DQC execution time and success probability. A plan with fewer remote operations may still incur long waits for probabilistic entanglement generation, additional swaps, or contention for communication memories. During these waits, stored qubits decohere. Researchers therefore need to evaluate execution time and success probability jointly across compilation, control, and network protocols.

We address this need by extending SeQUeNCe, the open-source R\&D 100 Award-winning discrete-event quantum network simulator~\cite{sequence}, with \dqc{}.
The module integrates network-aware compilation and controller--worker execution with SeQUeNCe's photon transmission, heralding, entanglement swapping, and memory models. This integration lets researchers test whether abstract compiler cost reductions translate into gains under simulated network dynamics. 
This is particularly important for compiler development: reducing a heuristic communication cost does not establish an application-level benefit unless entanglement generation, classical signaling, memory availability, and noise are evaluated together during execution.
Unlike an abstract Bell-pair operation, protocol-level simulation exposes the repeated attempts and classical signaling needed to establish remote entanglement. 
This distinction matters when comparing compilation strategies that place different demands on network paths and communication memories.

A central goal of this work is to provide the community with an easy-to-use open-source tool that advances DQC research by enabling researchers to realistically evaluate compilers, network protocols, and hardware architectures within a unified discrete-event simulation framework.
Our contributions are:
\begin{enumerate}
    \item Design and implementation of \dqc{}\footnote{\url{https://github.com/ssuresh27-byte/SeQUeNCe.DQC}} for full-stack DQC simulation in SeQUeNCe (\S\ref{sec:architecture}).
    
    \item Validation of \dqc{} by reproducing the results of a distributed Grover search experiment and comparing our simulated success probabilities with the reported 71\% success rate~\cite{grovers-nature25} (\S\ref{sec:grovers}).
    
    \item Simulation study on the impact of compiler-network interactions (compiler choice, network topology, and quantum memory) on the performance of DQC using \dqc{} in SeQUeNCe (\S\ref{sec:evaluation}). 
\end{enumerate}


\section{Design and Implementation}
\label{sec:architecture}

\begin{figure}[t]
    \centering
    \includegraphics[width=\linewidth]{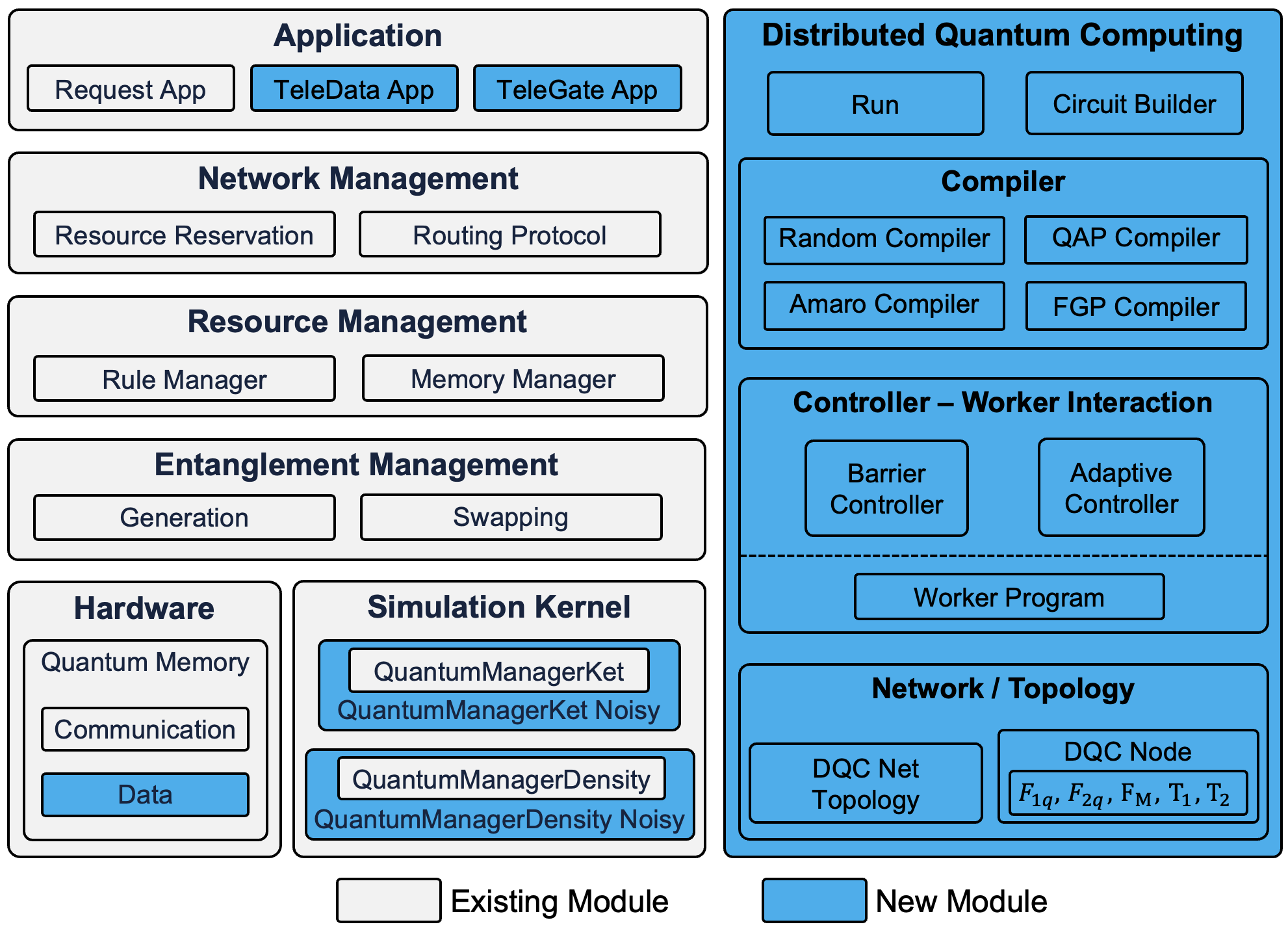}
    \vspace{-0.25in}
    \caption{Modular architecture of \dqc{}. Light-gray boxes represent existing SeQUeNCe modules; blue boxes highlight new modules added to enable full-stack DQC simulation.}
    \label{fig:sequence.dqc}
\end{figure}

\dqc{} is a SeQUeNCe top-level module that extends SeQUeNCe from a quantum \emph{network} simulator into a full-stack \emph{DQC} simulator.
It compiles a monolithic circuit onto a target topology and executes it over SeQUeNCe’s discrete event-driven simulator (Figure~\ref{fig:sequence.dqc}).
In this section, we introduce the \dqc{} design and implementation.

\subsection{Runtime Interface and Simulation Configuration}
Users interact with \dqc{} through a single high-level API entry point, \texttt{runtime.run} (see the top left of Figure~\ref{fig:pipeline}). 
Given a target \emph{circuit}, network \emph{topology}, and \emph{compiler}, \texttt{run()} automatically constructs the whole network topology~\texttt{DQCNetTopo}, instantiates worker processes, compiles the circuit, executes the discrete-event simulation timeline, and returns measurement outcomes across single or multi-shot runs. The circuit is generated by the function \texttt{CircuitBuilder.build()}$\to$\texttt{Circuit} which accepts OpenQASM, Qiskit, or QuTiP circuits. Two simulation backends are supported:
\begin{itemize}
    \item \textbf{Density-matrix backend}: Evaluates exact quantum states under completely positive and trace-preserving (CPTP) noise channels represented by Kraus operators. This backend has a high computational cost and is used for reproducing experimental results in simulations in~\S\ref{sec:grovers}.
    \item \textbf{Ket-vector backend}: Optimized for speed and higher qubit counts, modeling gate errors and pure dephasing derived from $T_1/T_2$ through per-shot random-Pauli trajectories, with $T_1$ relaxation approximated by stochastic resets. This backend drives the majority of our scaling evaluations in~\S\ref{sec:evaluation}.
\end{itemize}
Hardware error parameters—including single- and two-qubit gate infidelities, measurement errors, and $T_1/T_2$ coherence times—are specified per DQC Node, allowing for fine-grained modeling of heterogeneous hardware platforms under either backend formalism.

\subsection{Circuit Partitioning and Remote Operation Planning}
\begin{figure}[t]
\centering
\includegraphics[width=\columnwidth]{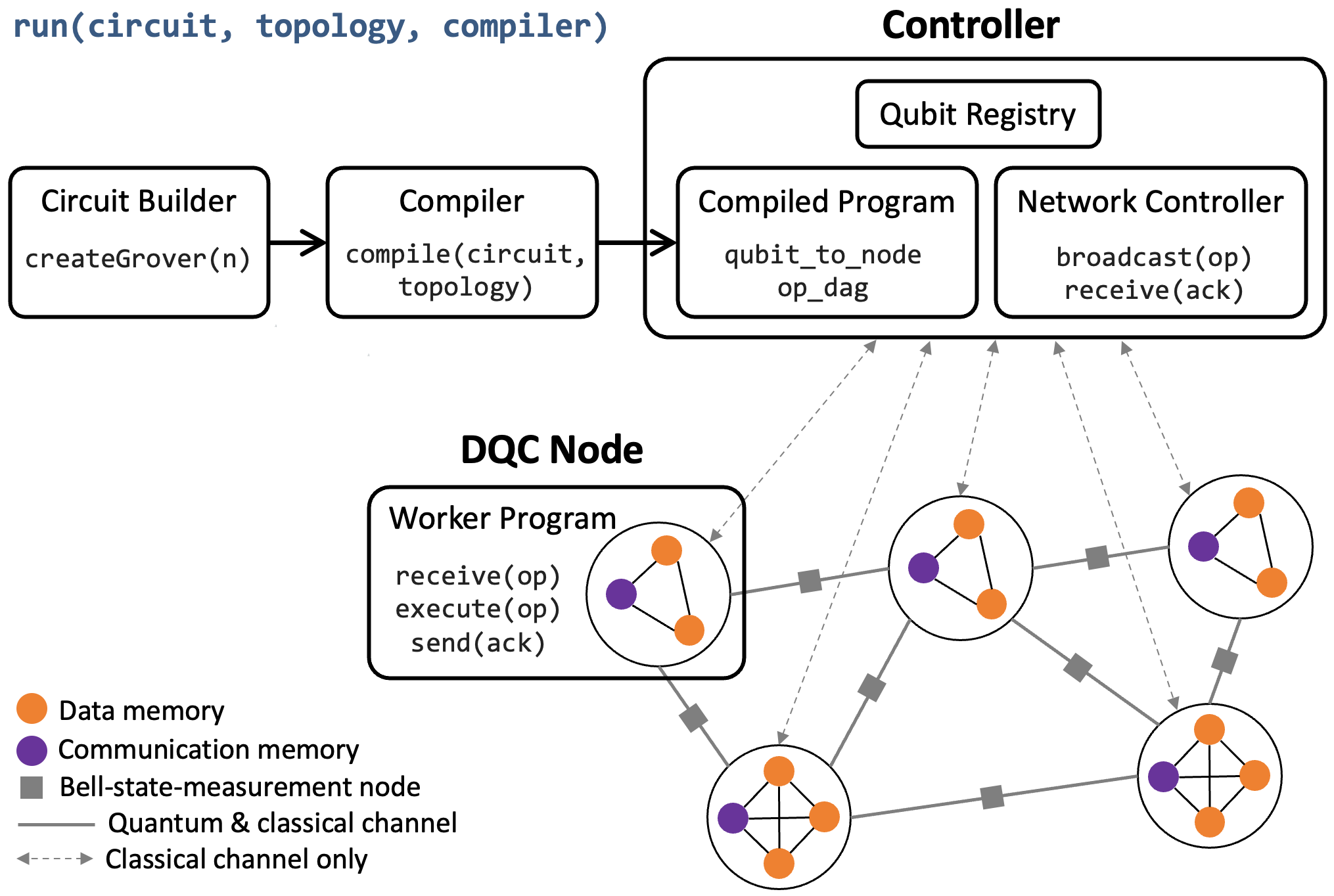}
\caption{\dqc{} execution workflow. The compiler produces a \texttt{CompiledProgram}; the controller tracks qubit locations and dispatches operations to per-node \texttt{WorkerPrograms}.}
\label{fig:pipeline}
\end{figure}

A compiler takes a quantum circuit and network topology as input and produces a \texttt{CompiledProgram} that the controller further executes (Figure~\ref{fig:pipeline}).
We assume that each QPU contains fully connected, equivalent qubits. Thus, no intra-QPU mapping optimization or routing is required, and compilation focuses on circuit partitioning across QPUs and remote operation planning.
The compiler makes two core decisions: circuit \emph{partitioning}, which assigns logical qubits to \texttt{DQCNode} instances subject to memory-capacity constraints, and \emph{remote operation planning}, which determines whether inter-node gates use \telegate{} or qubit relocations via \teledata{}.
The compiled program contains the initial qubit-to-node assignment and an operation dependency directed acyclic graph (DAG) comprising local gates, \telegate{} instructions, and \teledata{} instructions.

\dqc{} supports multiple compilation strategies. Built-in options include a quadratic assignment problem (\texttt{qap}) compiler that minimizes hop-weighted communication distance and a \texttt{random} partitioning baseline. Both retain fixed qubit-to-node assignments and implement all inter-node gates using \telegate{}. 
A time-sliced compiler based on FGP~\cite{fgp-2020} introduces \teledata{} moves when qubit relocation is advantageous, producing hybrid \telegate{}/\teledata{} execution plans.
The \texttt{CompiledProgram} provides a common interface that makes \dqc{} a \emph{compiler-agnostic simulation platform}: any compiler that produces a compatible \texttt{CompiledProgram} can use the same simulation runtime. External and custom compilers can therefore be evaluated using simulated physical network behavior, complementing their abstract cost estimates.~\S\ref{sec:evaluation} demonstrates this capability.

\subsection{Controller - Worker Interaction}

\textbf{Controller.}
The \texttt{Controller} is a classical node that connects to each \texttt{DQCNode} via classical channels.
It maintains a central \texttt{QubitRegistry} that tracks the dynamic mapping between logical circuit qubits and physical memory addresses across the network.
At initialization, a \texttt{Compiler} produces a \texttt{CompiledProgram} that sets up this registry. The controller then drives execution using one of two built-in control policies: a lock-step \emph{barrier} policy that broadcasts each circuit layer, waits for all acknowledgement (ACK) messages, and advances; or an \emph{adaptive} policy that dispatches individual operations as soon as their preceding dependencies resolve.
Because the controller tracks physical memory address mappings, it resolves logical target qubits to concrete physical locations prior to instruction dispatch. Whenever a \teledata{} operation completes, its ACK message updates the controller's registry with the qubit's new location.
Modeling classical transmission channels explicitly ensures that control-plane latencies and broadcast overheads are faithfully captured during simulation.

\textbf{DQC Node.} Each \texttt{DQCNode} serves as a worker that executes quantum operations dispatched by the controller.
It combines a QPU with quantum repeater functionality, following the server-centric architecture described in~\cite{benchmarking-QDC-QCE26}.
Each node has two distinct quantum memory resources: \emph{data memories} for computation and \emph{communication memories} for inter-node entanglement.
Communication qubits are allocated dynamically for entanglement-generation attempts to produce heralded Bell pairs and are recycled upon protocol completion.
A dedicated Bell-state-measurement node is placed in the middle of each  link connecting adjacent \texttt{DQCNode} instances.

Each \texttt{DQCNode} houses a local \texttt{WorkerProgram} that executes instructions dispatched by the central controller. The controller issues operations already resolved to concrete physical memory addresses on the target node.
The \texttt{WorkerProgram} executes local gates directly, or reserves network qubits to execute a nonlocal protocol. It returns an ACK upon operation completion; for nonlocal gates, this ACK is deferred until the underlying entanglement and gate protocols finish.

\subsection{Two Teleportation Primitives}
Nonlocal operations are delegated to SeQUeNCe's teleportation suite via a single teleportation application instance per node, exposing two primary primitives:
\begin{itemize}
    \item \telegate{} (\emph{gate teleportation}) executes a remote two-qubit gate (\texttt{cx} or \texttt{cz}) while leaving both data qubits on their respective home nodes.
    \item \teledata{} (\emph{state teleportation}) physically \emph{relocates} a data qubit to another node. This primitive supports two modes: moving a qubit into an empty memory slot, or performing an inter-node swap when target slots are occupied (requiring two concurrent \teledata{} operations). Because parallel links enable simultaneous entanglement generation, swap operations complete in roughly the same duration as single-qubit relocations.
\end{itemize}
These primitives represent a compilation trade-off: incurring a single \teledata{} relocation upfront can convert subsequent inter-node operations into fast local gates, reducing total \telegate{} overhead (\S\ref{sec:evaluation}).
Both primitives run atop SeQUeNCe's resource-management layer, meaning each required Bell pair may demand multiple probabilistic attempts, and multi-hop link interactions are enabled by entanglement swapping.

\section{Reproducing a Distributed Grover Experiment}
\label{sec:grovers}

To validate \dqc{} against experimental results, we reproduce the two-node distributed Grover search experiment of Main et al.~\cite{grovers-nature25} in simulation and compare success probabilities. In their setup, two photonically interconnected trapped-ion modules execute Grover's algorithm using teleported controlled-$Z$ gates.
Their reported benchmark is a $71\%$ average success probability across the four two-qubit marked states, achieved via a teleported-CZ average gate fidelity of $86.2\%$ and a remote Bell-pair fidelity of $96.89\%$.

\subsection{Setup and noise model}
\label{sec:noisemodel}
We construct the two-qubit Grover circuit and assign its two data qubits to separate modules connected by a single photonic channel. The CZ gates in the oracle and diffusion circuits are the only nonlocal operations, each implemented using \telegate{}.
We run the full \dqc{} stack under the density-matrix backend, injecting calibrated error channels from the experiment and estimating noisy-model success over $10{,}000$ shots per marked state ($40{,}000$ shots total).

We evaluate two noise configurations:
\begin{enumerate}
    \item \textbf{Noiseless control:} Isolates compilation and readout artifacts from hardware noise.
    \item \textbf{Detailed model:} Implements the experiment's full per-source error budget (Table~\ref{tab:budget})~\cite[Table~1]{grovers-nature25}. The six per-CZ noise sources act on the data qubits after each teleported CZ gate once communication qubits are traced out. Additionally, we apply local single-qubit gate errors ($1 \times 10^{-4}$) after every local $H$/$X$ gate and readout errors ($5.0 \times 10^{-3}$) at final readout.
\end{enumerate}

\begin{table}[t]
\centering
\caption{Implemented error budget ($A/B =$ Alice/Bob $=$ control/target)~\cite[Table~1]{grovers-nature25}.}
\label{tab:budget}
\footnotesize
\setlength{\tabcolsep}{3pt}
\resizebox{\linewidth}{!}{%
\begin{tabular}{@{}lccll@{}}
\toprule
Error source & A & B & Noise channel & Applied \\
\midrule
Raw entanglement & \multicolumn{2}{c}{$3.11\%$} & 2-qubit depol. & each CZ \\
Mixed-species gate & $2.4\%$ & $2.0\%$ & 1-qubit depol. & each CZ \\
QC decoherence ($T_2$) & $1.9\%$ & $1.8\%$ & 1-qubit deph.\ ($Z$) & each CZ \\
QX\,$\leftrightarrow$\,QC transfer & $0.76\%$ & $0.52\%$ & 1-qubit depol. & each CZ \\
Mid-circuit meas. & $0.091\%$ & $0.122\%$ & 1-qubit flip ($X$) & each CZ \\
In-CZ rotations & $0.016\%$ & $0.015\%$ & 1-qubit depol. & each CZ \\
\midrule
Local gate ($H$/$X$) & \multicolumn{2}{c}{$1{\times}10^{-4}$} & 1-qubit depol. & every local \\
Readout & \multicolumn{2}{c}{$5.0{\times}10^{-3}$} & bit-flip & final meas. \\
\bottomrule
\end{tabular}%
}
\end{table}

\subsection{Results}
\begin{figure}[t]
\centering
\includegraphics[width=0.92\linewidth]{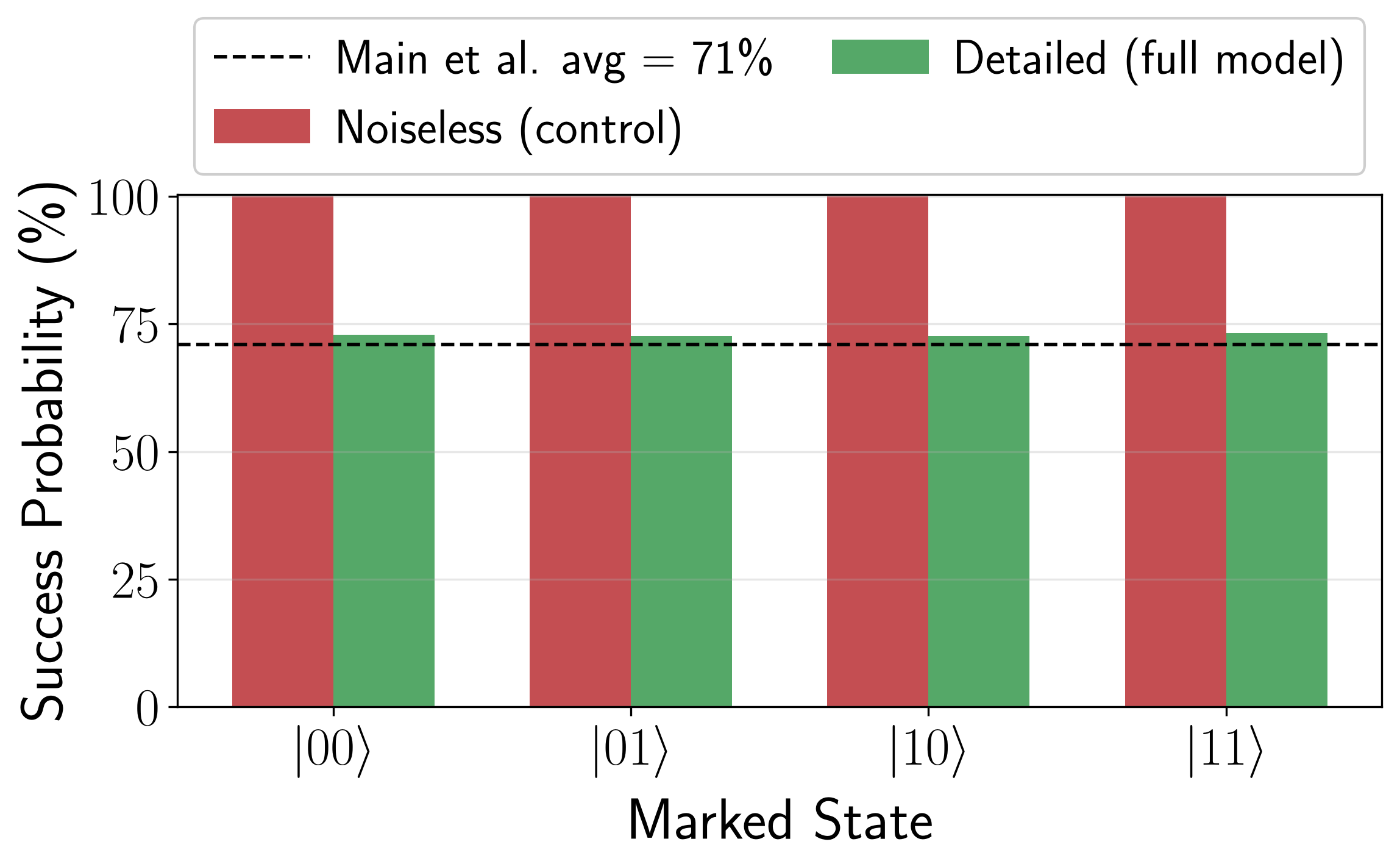}
\vspace{-0.1in}
\caption{Two-node distributed Grover validation. Per-marked-state success probability for the noiseless baseline ($100\%$) and the full detailed error model ($72.9\%$ avg; $10{,}000$ shots per marked state). The dashed line represents the measured $71\%$ average from Main et al.~\cite{grovers-nature25}.}
\label{fig:grover}
\end{figure}

Figure~\ref{fig:grover} compares per-marked-state success probabilities for the two noise configurations with the experimental average.
The noiseless baseline achieves a $100\%$ success rate across all four states, confirming that the teleported-CZ execution and readout routines function correctly end-to-end.
When noise channels are enabled, the simulated success probabilities approach the experimental benchmark: evaluated over $10{,}000$ shots per state, the full detailed model achieves $72.9\%$ on average (per-state: $\lvert00\rangle 72.9\%$, $\lvert01\rangle 72.7\%$, $\lvert10\rangle 72.7\%$, $\lvert11\rangle 73.2\%$).
The per-state success probabilities span only $0.5$ percentage points, showing similar observed performance across the four marked states.

\para{Findings.}
Relative to the noiseless baseline, modeled noise reduces success probability by approximately $27.1$ percentage points.
The detailed model yields a success probability $1.8\%$ above the experimental average from \cite{grovers-nature25}. This difference may partly reflect an underestimate of the experimental gate errors: Main et al.~\cite{grovers-nature25} report that the measured teleported-CZ error exceeds the component error-budget prediction and attribute this discrepancy to calibration drift.

\section{Evaluation on Compiler, Network \& Memory}
\label{sec:evaluation}

Following validation against the two-node Grover experiment, we use \dqc{} to study how compiler strategy, network topology, and memory-capacity distribution affect execution time and success probability for larger Grover circuits.

Throughout this section, Grover's success probability is estimated from 1,000 distributed circuit executions per configuration. We first evaluate simulator scaling and noise sensitivity, then compare architectural choices using the ket-vector backend at $z=0.95$ (Table~\ref{tab:zsweep}). Table~\ref{tab:configs} summarizes the experimental configurations.
For an $n$-qubit Grover search, the circuit builder uses $n-2$ ancilla qubits to decompose multi-controlled operations, giving $2n-2$ circuit qubits in total, all stored in data memories, excluding communication qubits.
In Table~\ref{tab:configs}, Data and Comm denote the numbers of data and communication qubits per node. 
Throughout this section, links are 5 km long, light propagates through fiber at $2\times10^8\,\mathrm{m/s}$, and memory and detector efficiencies are both 0.9. Local gates are assumed to take $1 \mu$s (typical for neutral atoms). 


\begin{table}[ht]
\centering
\caption{Experimental configurations.}
\label{tab:configs}
\resizebox{\linewidth}{!}{%
\begin{tabular}{@{}lllcc@{}}
\toprule
Evaluation Target & $n$ & Topology & Data & Comm\\
\midrule
Scaling Limits (Table~\ref{tab:scaling}) & $2$--$10$ & Star ($2n-2$ nodes)        & $1$ & $8$--$34$ \\
Noise Sensitivity (Fig.~\ref{fig:noise}) & $6$  & 2-node star             & $5$ & $8$ \\
Compilers (Fig.~\ref{fig:compilers}) & $4$   & 2-node link             & $3$ & $8$  \\
Topology (Fig.~\ref{fig:topocap}a)   & $6$  & Star / Grid / Caveman   & $1$ & $8$ \\
Capacity (Fig.~\ref{fig:topocap}b)   & $7$  & 4-node line             & Varied (12 total) & $8$\\
\bottomrule
\end{tabular}%
}
\end{table}

\subsection{Simulator Scaling and Noise Sensitivity}
\label{sec:calibration}

\textbf{Scaling constraints.}
To examine the computational cost of full-state simulation, we evaluate distributed Grover on a $(2n-2)$-node star network (sweeping $n$ from $2$ to $10$ data qubits) where every two-qubit gate requires an inter-node \telegate{}. 
The hub node in the star network is provisioned with $\max(8, 4n{-}6)$ communication qubits to accommodate increasing telegate concurrency as circuit width grows.
As shown in Table~\ref{tab:scaling}, simulated Grover search circuit execution time increases from $10.8$\,ms to $7.4$\,s ($\sim\!690\times$), while simulator host wall-clock time grows exponentially.
The largest evaluated circuit contains ten search qubits and eight ancillas and requires approximately $71$ minutes of wall-clock time on an Apple M3 Pro chip, illustrating the computational cost of full-state discrete-event DQC simulation.

\begin{figure}[t]
\centering
\includegraphics[width=0.7\linewidth]{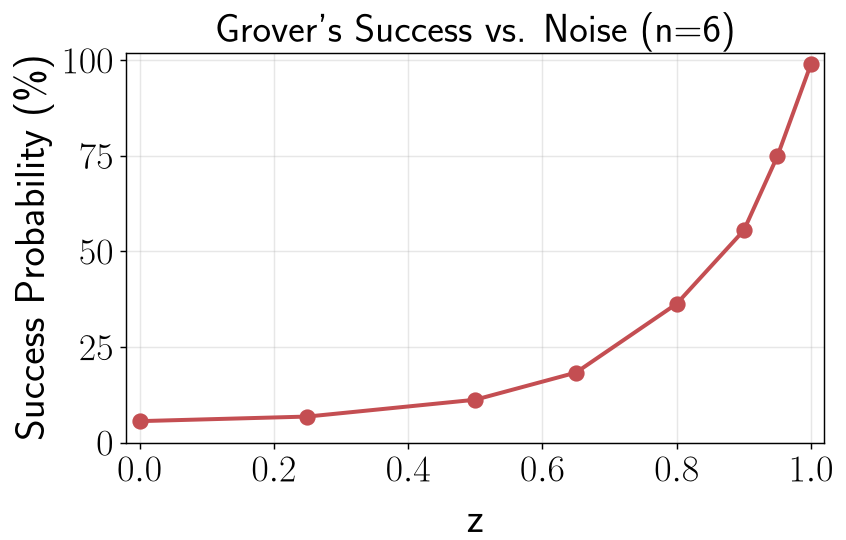}
\vspace{-0.1in}
\caption{Distributed Grover ($n{=}6$) success probability versus hardware quality $z$ (Table~\ref{tab:zsweep}), linking physical device noise directly to algorithmic outcome.}
\vspace{-0.1in}
\label{fig:noise}
\end{figure}

\begin{table}[t]
\centering
\caption{Distributed Grover scaling vs.\ problem size ($n$). Wall time is measured for a single circuit execution.}
\label{tab:scaling}
\scriptsize
\setlength{\tabcolsep}{4.5pt}
\begin{tabular}{@{}rrrrrrrr@{}}
\toprule
$n$ & Nodes & Qubits & Iter. & \telegate{} & \shortstack{Local\\gates} & \shortstack{Exec. time\\(ms)} & \shortstack{Wall time\\(s)} \\
\midrule
2  & 2  & 2  & 1  & 2    & 14   & 10.8   & 0.0    \\
3  & 4  & 4  & 2  & 52   & 107  & 70.2   & 0.1    \\
4  & 6  & 6  & 3  & 150  & 280  & 207.4  & 0.4    \\
5  & 8  & 8  & 4  & 296  & 533  & 423.3  & 1.1    \\
6  & 10 & 10 & 6  & 588  & 1038 & 887.4  & 3.5    \\
7  & 12 & 12 & 8  & 976  & 1703 & 1467.7 & 13.2   \\
8  & 14 & 14 & 12 & 1752 & 3032 & 2630.4 & 71.5   \\
9  & 16 & 16 & 17 & 2890 & 4973 & 4382.5 & 1249.3 \\
10 & 18 & 18 & 25 & 4850 & 8310 & 7424.7 & 4236.3 \\
\bottomrule
\end{tabular}
\end{table}
\textbf{Cross-layer noise sensitivity.}
To examine noise sensitivity, we use a coordinated hardware quality parameter $z \in [0, 1]$ proposed in~\cite{qre-qce26} that concurrently scales device-level gate fidelities, measurement errors, Bell-pair fidelity ($F_{\text{phys}}$), and dephasing times ($T_2$) (Table~\ref{tab:zsweep}).
Sweeping $z$ on an $n=6$ (10-qubit) two-node star illustrates how lower-level errors compound across teleported gates (Figure~\ref{fig:noise}). We select $z=0.95$ as an illustrative nonideal operating point for subsequent architectural comparisons. At this setting, the baseline success probability is $74.9\%$, allowing architectural effects to be examined in the presence of hardware noise.

\begin{table}[t]
\centering
\caption{Coordinated hardware quality parameter $z$ ($T_1{=}100$~s fixed).}
\label{tab:zsweep}
\scriptsize
\setlength{\tabcolsep}{3.2pt}
\begin{tabular}{@{}c|ccccc@{}}
\toprule
$z$ & $F_{1q}$ & $F_{2q}$ & $F_{m}$  & $F_{\text{phys}}$ & $T_2$ (s) \\
\midrule
$0.00$ & $0.99900$ & $0.99910$ & $0.99600$ & $0.96500$ & $2.00$ \\
$0.25$ & $0.99925$ & $0.99933$ & $0.99700$ & $0.97375$ & $2.67$ \\
$0.50$ & $0.99950$ & $0.99955$ & $0.99800$ & $0.98250$ & $4.01$ \\
$0.65$ & $0.99965$ & $0.99969$ & $0.99860$ & $0.98775$ & $5.73$ \\
$0.80$ & $0.99980$ & $0.99982$ & $0.99920$ & $0.99300$ & $10.03$ \\
$0.90$ & $0.99990$ & $0.99991$ & $0.99960$ & $0.99650$ & $20.07$ \\
$0.95$ & $0.99995$ & $0.99996$ & $0.99980$ & $0.99825$ & $40.14$ \\
$1.00$ & $1.00000$ & $1.00000$ & $1.00000$ & $1.00000$ & $199.99$ \\
\bottomrule
\end{tabular}
\end{table}

\subsection{Compiler choice}
\label{sec:compilers}
\begin{figure}[t]
\centering
\includegraphics[width=\linewidth]{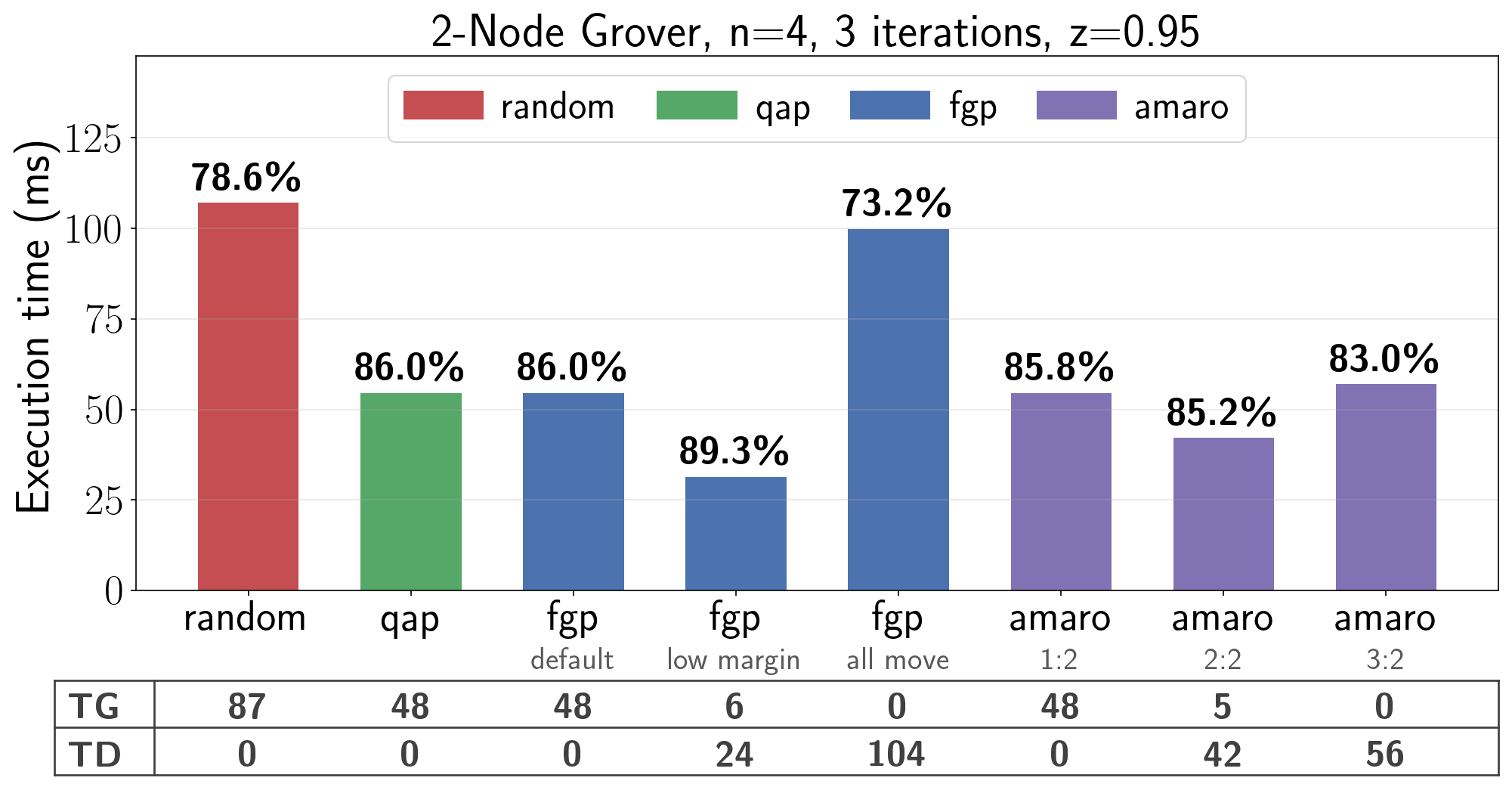}
\vspace{-0.25in}
\caption{Execution time and Grover's success probability of eight compilations of the same 2-node Grover ($n{=}4$). Bars annotate success at $z{=}0.95$.}
\label{fig:compilers}
\end{figure}

\para{Setup.}
Two nodes are joined by a single physical link, each with a data-memory capacity of three.
The workload is optimal-iteration (three-iteration) distributed 4-qubit Grover (six qubits total, including the two ancillas of the CX-decomposed oracle), with marked state $\lvert1111\rangle$.
Using four compilers, we compile the workload eight ways: the \texttt{random} and \texttt{qap} partition baselines, our FGP time-sliced compiler at three operating points, and one external compiler to demonstrate framework flexibility.
Because \dqc{} is compiler-agnostic, its \texttt{run} function accepts any object implementing \texttt{compile(circuit, topology, seed)}$\to$\texttt{CompiledProgram}. Users can supply custom circuit partitioning and remote operation planning strategies or external tool outputs, which share the same simulation runtime.
We demonstrate this capability with Amaro~\cite{amaro-2025}, a compiler generator synthesizing qubit mapping-and-routing compilers from high-level device specifications.
Our specification (\texttt{dqc\_hybrid.qmrl}) models each QPU as an internal clique. Inter-node interactions use either an in-place \telegate{} at \texttt{link\_cost} or a \teledata{} relocation at \texttt{teleport\_cost}. Amaro generates a solver that minimizes the specified cost of circuit partitioning and remote operation planning.
An \texttt{AmaroCompiler} adapter converts the circuit to OpenQASM and the network to an architecture graph, then translates the solver's per-step qubit assignments into a \texttt{CompiledProgram}. This allows compiler outputs to be compared using simulated execution time and success probability, alongside their abstract costs.
We evaluate \telegate{}:\teledata{} cost ratios of $1{:}2$, $2{:}2$, and $3{:}2$ (Figure~\ref{fig:compilers}). These ratios specify the relative costs assigned to one \telegate{} operation and one \teledata{} move in the compiler's objective, rather than their simulated execution times.

\para{Findings.}
The \texttt{random} baseline scatters qubits and requires $87$ \telegate{}s on average ($107$\,ms), whereas \texttt{qap} reduces this to $48$ \telegate{}s ($54.6$\,ms).
At default operating points, both FGP and Amaro adopt pure-\telegate{} strategies on Grover's uniform interaction pattern.
When incentivized, however, both leverage the \telegate{}/\teledata{} trade-off: lowering FGP's relocation margin yields a hybrid plan with $6$ \telegate{}s and $24$ \teledata{} moves ($31.4$\,ms)—the fastest plan evaluated, achieving a $42\%$ runtime reduction over \texttt{qap}. 
Lowering the margin even further results in a pure-\teledata{} schedule ($104$ moves; $99.9$\,ms). Increasing Amaro's assigned \telegate{} cost relative to \teledata{} produces a hybrid plan ($5$ \telegate{}s and $42$ \teledata{} moves; $42.1$\,ms). At a cost ratio of $3{:}2$, Amaro produces a pure-\teledata{} schedule with $56$ moves, taking $56.9$\,ms---slightly longer than the pure-\telegate{} plan's $54.6$\,ms.

Success rate generally decreases with increased number of \telegate{} and \teledata{} operations, because more network operations result in extra entanglement overhead. 
The random and FGP all-move strategies produced success rates of $78.6\%$ and $73.2\%$, respectively, whereas the FGP low-margin compiler produced the highest success rate of $89.3\%$. 

\subsection{Network Topology}
\label{sec:topo}
\para{Setup.}
We evaluate the FGP compiler across three $10$-node networks with distinct edge structures: a \emph{star} (one hub, nine spokes), a \emph{grid} ($2{\times}5$ mesh), and a \emph{caveman} graph (two $5$-cliques joined by a single bridge).
Because each node has a capacity of $1$ (holding $10$ total qubits for an $n = 6$ distributed Grover circuit), every two-qubit gate is inherently an inter-node \telegate{}. We measure success under $z = 0.95$ noise.

\para{Findings.}
FGP produces $588$ \telegate{} operations and no \teledata{} moves on all three topologies (Figure~\ref{fig:topocap}a). With one data-memory slot per node, relocation cannot make gates local.
Instead, performance is governed by the \emph{hop distance} of \telegate{} operations. Single-hop \telegate{}s require one heralded Bell pair, whereas two-hop \telegate{}s require \emph{two} pairs connected by an entanglement swap, adding latency and error overhead. The star topology routes spoke-to-spoke gates through the hub, resulting in $396$ two-hop \telegate{}s; the grid has $192$, and the caveman has only $48$ because interactions concentrate inside cliques. Total Bell-pair consumption drops from $984$ (star) to $780$ (grid) to $636$ (caveman).

Execution times are $887.4$\,ms, $818.1$\,ms, and $733.6$\,ms for star, grid, and caveman, respectively. Caveman has the lowest Bell-pair requirement and the highest observed success probability ($27.1\%$). The grid follows ($21.2\%$), while the star---despite the identical \telegate{} count---has the \emph{lowest} success ($15.0\%$): its $396$ two-hop \telegate{}s each consume two heralded pairs across the 5\,km links, so it accumulates the most entanglement noise and idle dephasing.

\subsection{Memory-Capacity Distribution}
\label{sec:capacity}
\para{Setup.}
We evaluate a fixed four-node line network ($m_0$--$m_1$--$m_2$--$m_3$) with a constant total data capacity of twelve slots. We vary slot allocation across nodes: balanced ($[3,3,3,3]$), mid-concentrated ($[6,3,2,1]$), and highly skewed ($[9,1,1,1]$). The workload is $n = 7$ optimal-iteration distributed Grover ($12$ total qubits: $7$ data $+$ $5$ ancillas), completely filling available data memory. We employ FGP with a reduced relocation margin to enable hybrid \telegate{} and \teledata{} execution.

\begin{figure}[t]
\centering
\includegraphics[width=\linewidth]{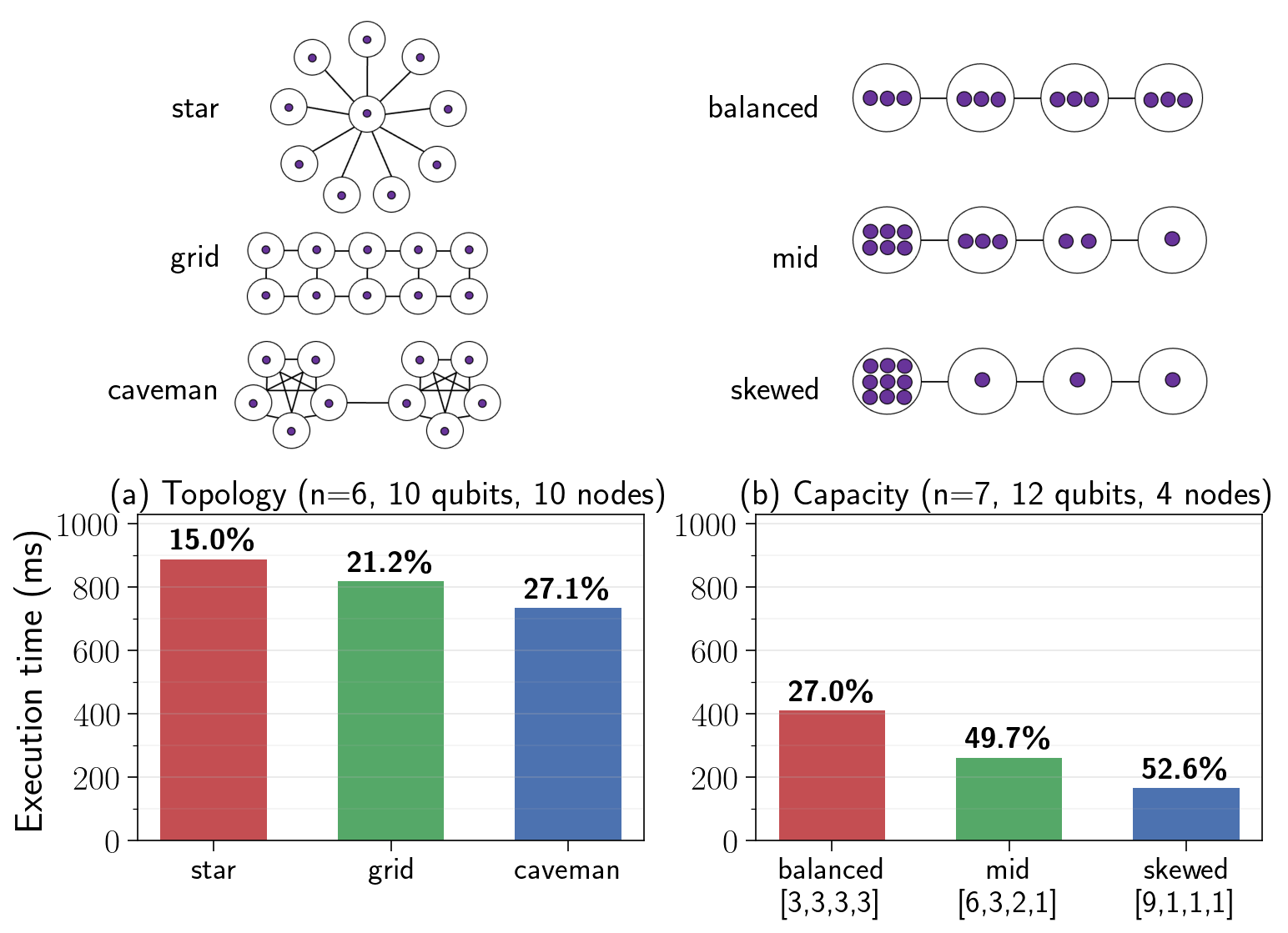}
\vspace{-0.25in}
\caption{Same FGP compiler; the network drives execution time and success probability.
(a) Three $10$-node topologies at capacity $1$: identical \telegate{} count ($588$), yet the star's hub serializes traffic and runs $\sim\!20\%$ slower.
(b) Fixed four-node line, fixed total memory ($12$ slots): concentrating capacity co-locates interacting qubits.}
\label{fig:topocap}
\end{figure}

\para{Findings.}
Reallocating memory capacity across nodes significantly alters execution time and success probability (Fig.~\ref{fig:topocap}b).
Under the selected FGP policy, the balanced layout ($[3,3,3,3]$) produces $369$ network operations ($17$ \telegate{}s and $352$ \teledata{} moves). This high network operation count yields the longest runtime ($410.7$\,ms) and lowest success probability ($27.0\%$) under $z = 0.95$ noise.
In contrast, concentrating memory capacity creates a hub co-locating frequently interacting qubits. The mid-concentrated layout reduces total network operations to $275$ ($143$ \telegate{}s $+$ $132$ moves), lowering runtime to $262.0$\,ms and raising success probability to $49.7\%$. The highly skewed layout ($[9,1,1,1]$) achieves the best performance, dropping operations to $145$ ($17$ \telegate{}s $+$ $128$ moves), reducing execution time to $165.5$\,ms, and achieving a $52.6\%$ success rate.
In these evaluations, topology changes communication paths, whereas memory allocation changes the mix and number of remote operations.

\section{Conclusion and Future Work}
\label{sec:conclusion}

We presented \dqc{}, extending the open-source SeQUeNCe simulator with circuit compilation and execution control integrated with heralded entanglement protocols for full-stack DQC evaluation.
We validated the framework by reproducing a two-node trapped-ion Grover experiment in simulation, obtaining $72.9\%$ success under the detailed noise model versus the measured $71\%$.
Hybrid \telegate{}/\teledata{} plannings with selective qubit relocation outperform both gate teleportation only and aggressive state movement in the evaluated workload.
This finding motivates further research into compilation algorithms that jointly optimize remote gates and qubit relocation while accounting for network topology and communication costs.
Our experiments also show that DQC performance depends on communication paths and memory distribution, even when the number of remote gates or total memory capacity is fixed.

Future work will extend \dqc{} to switch-centric quantum data center networks~\cite{benchmarking-QDC-QCE26}.
Our wide-area simulations suggest that small-area quantum data centers may better support DQC by reducing communication overhead.
We will also study scheduling and resource contention when multiple distributed circuit requests overlap.
Finally, we will extend \dqc{} beyond its current assumptions of full intra-QPU connectivity and uncorrected noise, incorporating realistic internal topologies and fault-tolerance models.

\bibliographystyle{IEEEtran}
\bibliography{bib}

\end{document}